\documentclass[11pt,twoside,fleqn]{article}

\usepackage{latexsym} 
\usepackage{amsfonts} 
\usepackage{epsf} 
\usepackage{graphicx} 
\usepackage{epsfig}

\newcommand{\ket}[1]{\left|#1\right>}

\newcommand{\f}[1]{\mbox{\boldmath$#1$}}
\newcommand{\fk}[1]{\mbox{\boldmath$\scriptstyle#1$}}
\newcommand{\vau}{\mbox{\boldmath$v$}}
\newcommand{\na}{\mbox{\boldmath$\nabla$}}
\newcommand{\bea}{\begin{eqnarray}}
\newcommand{\ea}{\end{eqnarray}}
\newcommand{\eea}{\end{eqnarray}}
\newcommand{\ord}{{\cal O}}

 \def\beq{\begin{equation}}
 \def\eeq{\end{equation}}
 \def\bea{\begin{eqnarray}}
 \def\eea{\end{eqnarray}}

\begin{document} 

\title{\bf Refractive index perturbations -- Unruh effect, 
Hawking radiation or dynamical Casimir effect?}

\author{Ralf Sch\"utzhold\\
Fakult\"at f\"ur Physik, Universit\"at Duisburg-Essen, 
Lotharstrasse 1, 47057 Duisburg, Germany\\
{\tt ralf.schuetzhold@uni-due.de}
}

\maketitle

%%%%%%%%%%%%%%%%%%%%%%%%%%%%%%%%%%%%%%%%%%%%%%%%%%%%%%%%%%%%%%%%%%%%%%%%%%%%%%%%%%%%%%%%%%%%%%%
\section{Introduction}
%%%%%%%%%%%%%%%%%%%%%%%%%%%%%%%%%%%%%%%%%%%%%%%%%%%%%%%%%%%%%%%%%%%%%%%%%%%%%%%%%%%%%%%%%%%%%%%

Motivated by recent experimental progress to manipulate the refractive index of dielectric 
materials by strong laser beams, we study some aspects of the quantum radiation created by 
such refractive index perturbations. 
The Kerr effect in non-linear dielectrics describes a change of the refractive index $n$ 
due to electromagnetic radiation with the intensity $I$ 
\bea
n=n_0+n_2I
\,,
\ea
where $n_0$ is the unperturbed index of refraction and $n_2$ is the Kerr coefficient. 
For example, for fused silica, it is  $n_2=3\times10^{-16}\,\rm W^{-1}cm^2$.
Since the intensity $I$ will be far below $1/n_2$, the change $\delta n$ in the  
refractive index $n$ will be small.
However, with a space-time dependent intensity $I(t,\f{r})$, we can generate 
a varying refractive index perturbation 
\bea
n(t,\f{r})=n_0+\delta n(t,\f{r})
\,,\quad
\delta n\ll1
\,.
\ea
Such a perturbation is able to create photon pairs out of the vacuum -- 
more precisely, the ground state of the quantized electromagnetic field
in the dielectric medium. 
In order to investigate this effect, we will make the following simplifications: 
\begin{itemize}
\item{We assume zero temperature and thus start in the ground state.}
\item{We omit the photon polarizations and consider scalar fields.}
\item{We neglect the dispersion $n(\omega)$ of the medium.} 
\item{Since $\delta n\ll1$ is small, we use perturbation theory.} 
\end{itemize}
The first two assumptions are not critical and not difficult to fix.
However, taking into account the dispersion $n(\omega)$ of the medium 
is rather non-trivial and calculations beyond perturbation theory are 
also far more involved in general. 

%%%%%%%%%%%%%%%%%%%%%%%%%%%%%%%%%%%%%%%%%%%%%%%%%%%%%%%%%%%%%%%%%%%%%%%%%%%%%%%%%%%%%%%%%%%%%%%
\section{Perturbation Theory} 
%%%%%%%%%%%%%%%%%%%%%%%%%%%%%%%%%%%%%%%%%%%%%%%%%%%%%%%%%%%%%%%%%%%%%%%%%%%%%%%%%%%%%%%%%%%%%%%

Given the aforementioned assumptions, we may start from the Lagrangian 
($\hbar=c_0=1$)
\bea
{\cal L}
=
\frac12\left(\varepsilon E^2-B^2\right)
=
\frac12\left([n_0+\delta n]^2E^2-B^2\right)
\,,
\ea
where $\varepsilon$ is the dielectric permittivity of the medium and 
$E$ and $B$ are the microscopic electric and magnetic fields, respectively. 
Now standard quantization procedure yields the Hamiltonian 
\bea
\hat H
=
\frac12\int d^3r\left(\frac{\hat D^2}{[n_0+\delta n]^2}+\hat B^2\right)
=
\hat H_0
+
\hat H_1
\,,
\ea
where $D=\varepsilon E$ is the electric displacement field including the 
polarization $P$ of the medium $D=E+P$.
Using $\delta n\ll1$, we expand $\hat H$ into powers of $\delta n$ such 
that the lowest order $\hat H_0$ is the undisturbed Hamiltonian and 
the rest $\hat H_1$ contains the effect of the refractive index perturbation.
Then we may use usual time-dependent perturbation theory in the interaction 
picture to calculate the final quantum state $\ket{\Psi_{\rm out}}$ starting
from the initial vacuum state $\ket{0}$ 
\bea
\ket{\Psi_{\rm out}}=\ket{0}-i\int dt\,\hat H_1(t)\ket{0}+\ord(\delta n^2)
\,.
\ea
Inserting the expansion into creation and annihilation operators for 
the quantized electromagnetic field, we find the production of photon pairs
(to lowest order in $\delta n$)
\bea
\ket{\Psi_{\rm out}}
=
\ket{0}+\sum\limits_{\fk{k},\fk{k'}}
{\cal A}_{\fk{k},\fk{k'}}\ket{\f{k},\f{k'}}
+\ord(\delta n^2)
\,,
\ea
with the two-photon amplitude/probability (see, e.g., \cite{External} and \cite{Birula})
\bea
\left|{\cal A}_{\fk{k},\fk{k'}}\right|^2
=
\frac{\omega_k\omega_k'}{n_0^6}\,
\left|
\widetilde{\delta n}(\omega_k+\omega_k',\f{k}+\f{k'})
\right|^2
\,.
\ea
Here $\f{k}$ and $\f{k'}$ are the wave-numbers of the created photon pair 
and $\omega_k$ $\omega_k'$ their frequencies. 

%%%%%%%%%%%%%%%%%%%%%%%%%%%%%%%%%%%%%%%%%%%%%%%%%%%%%%%%%%%%%%%%%%%%%%%%%%%%%%%%%%%%%%%%%%%%%%%
\section{Single One-Parameter Pulse} 
%%%%%%%%%%%%%%%%%%%%%%%%%%%%%%%%%%%%%%%%%%%%%%%%%%%%%%%%%%%%%%%%%%%%%%%%%%%%%%%%%%%%%%%%%%%%%%%

As a first example, let us consider a simple pulse whose spatial 
and temporal extent are given by the same parameter $\Omega$, i.e., 
\bea
\delta n(t,\f{r})=\delta\bar n\,f(\Omega t,\Omega \f{r}/c)
\,,
\ea
with a well behaved function $f=\ord(1)$ (e.g., Gaussian).
Then it turns out that the total emission probability is independent of $\Omega$ 
\bea
P=
\sum\limits_{\fk{k},\fk{k'}}
\left|{\cal A}_{\fk{k},\fk{k'}}\right|^2
=
\ord(\delta\bar n^2)
\ll1
\,,
\ea
where the precise pre-factor in front of $\delta\bar n^2$ depends on the 
precise pulse shape.  
The typical photon energy, however, is given by  
\bea
E=
\frac1P
\sum\limits_{\fk{k},\fk{k'}}
\left|{\cal A}_{\fk{k},\fk{k'}}\right|^2
\omega_k
=
\ord(\Omega)
\,.
\ea
Nevertheless, the total probability is bound to be small.
For an experimental verification, therefore, a pulse with more 
than one parameter might be more useful.  

%%%%%%%%%%%%%%%%%%%%%%%%%%%%%%%%%%%%%%%%%%%%%%%%%%%%%%%%%%%%%%%%%%%%%%%%%%%%%%%%%%%%%%%%%%%%%%%
\section{Single Two-Parameter Pulse}
%%%%%%%%%%%%%%%%%%%%%%%%%%%%%%%%%%%%%%%%%%%%%%%%%%%%%%%%%%%%%%%%%%%%%%%%%%%%%%%%%%%%%%%%%%%%%%%

Thus, as a second example, let us assume that the spatial and temporal scales of the 
pulse are given by two different parameters, $\Omega_1$ and $\Omega_2$
\bea
\delta n(t,\f{r})=\delta\bar n\,f(\Omega_1 t,\Omega_2 \f{r}/c)
\,.
\ea
If $\Omega_1$ and $\Omega_2$ are of the same order, we basically reproduce the previous case. 
Thus, let us consider the two limiting cases $\Omega_1\ll\Omega_2$ and $\Omega_1\gg\Omega_2$. 
\begin{itemize}
\item For $\Omega_1\ll\Omega_2$, the spatial extend is very small and we have 
a ``point-like pulse''. 
\item For $\Omega_1\gg\Omega_2$, the pulse is almost spatially homogeneous and thus 
the situation is very similar to ``cosmological particle creation''.  
\end{itemize}
In the first case (``point-like pulse''), one can derive a closed expression for the 
average energy emitted by the refractive index perturbation \cite{External}
\bea
\langle E\rangle 
\propto
%\frac{n_0^6}{105(2\pi)^3}
\int dt\left(
\frac{d^4}{dt^4}
\left[\int d^3r\,\delta n(t,\f{r})\right]
\right)^2
=
\ord\left(\frac{\Omega_1^7}{\Omega_2^6}\,\delta\bar n^2\right)
\,.
\ea
However, since the typical photon energy is set by the temporal 
frequency $\Omega_1$ and not by the spatial scale $\Omega_2$, 
the total probability is extremely small 
\bea
E=\ord(\Omega_1)
\,\leadsto\,
P=
\ord\left(\frac{\Omega_1^6}{\Omega_2^6}\,\delta\bar n^2\right)
\ll1
\,.
\ea
Thus this effect is also probably very hard to detect. 

%%%%%%%%%%%%%%%%%%%%%%%%%%%%%%%%%%%%%%%%%%%%%%%%%%%%%%%%%%%%%%%%%%%%%%%%%%%%%%%%%%%%%%%%%%%%%%%
\section{Cosmological Particle Creation}
%%%%%%%%%%%%%%%%%%%%%%%%%%%%%%%%%%%%%%%%%%%%%%%%%%%%%%%%%%%%%%%%%%%%%%%%%%%%%%%%%%%%%%%%%%%%%%%

In the opposite limit $\Omega_1\gg\Omega_2$, we may approximate the pulse with 
\bea
\delta n(t,\f{r})=\delta\bar n\,f(\Omega_1 t,\Omega_2 \f{r}/c)
\ea
by a effectively spatially homogeneous profile $\delta n(t,\f{r})\approx\delta n(t)$ 
and thus the Lagrangian reads 
\bea
{\cal L}
\approx
\frac12\left(\varepsilon(t)E^2-B^2\right)
=
\frac12\left(n^2(t)E^2-B^2\right)
\,.
\ea
This is completely analogous to an expanding/contracting universe where the scale factor 
$a^2(t)$ corresponds to the refractive index $n(t)$. 
Hence we can immediately apply the results known for this situation, 
see, e.g., \cite{Birrell} and references therein. 
\begin{itemize}
\item Photons are created in pairs with nearly opposite momenta $\f{k'}\approx-\f{k}$.
\item Thus, they are in an entangled (squeezed) state.  
\item The typical photon energy scales with $E=\ord(\Omega_1)$.  
\item The total probability is enlarged by a volume enhancement factor $\Omega_1^3/\Omega_2^3$
\bea
P=
\ord\left(\frac{\Omega_1^3}{\Omega_2^3}\,\delta\bar n^2\right)
\,.
\ea
\end{itemize}
In view of this enhancement, this effect might well be observable. 

%%%%%%%%%%%%%%%%%%%%%%%%%%%%%%%%%%%%%%%%%%%%%%%%%%%%%%%%%%%%%%%%%%%%%%%%%%%%%%%%%%%%%%%%%%%%%%%
\section{Mixed Case}
%%%%%%%%%%%%%%%%%%%%%%%%%%%%%%%%%%%%%%%%%%%%%%%%%%%%%%%%%%%%%%%%%%%%%%%%%%%%%%%%%%%%%%%%%%%%%%%

For completeness, let us briefly discuss the scenario with three different parameters
\bea
\delta n(t,\f{r})=\delta\bar n\,f(\Omega_1 t,\Omega_2 x/c,\Omega_3 y/c,\Omega_3 z/c) 
\,,
\ea
where one spatial scale, say $\Omega_2$, is much smaller than the temporal frequency
$\Omega_1$ while the other, $\Omega_3$, is much larger -- corresponding to a long 
``needle-like pulse'' pointing in $x$-direction
\bea
\Omega_2 \ll \Omega_1 \ll \Omega_3
\,.
\ea
Again, the typical photon energy is set by the temporal frequency $\Omega_1$ 
(which is the generic case). 
The total probability is enhanced by a length factor $\Omega_1/\Omega_2$
but suppressed by the small cross section area $\propto1/\Omega_3^2$ 
of the ``needle'' 
\bea
P=\ord\left(\frac{\Omega_1}{\Omega_2}\,\frac{\Omega_1^4}{\Omega_3^4}\,\delta\bar n^2\right)
\,.
\ea
%

%%%%%%%%%%%%%%%%%%%%%%%%%%%%%%%%%%%%%%%%%%%%%%%%%%%%%%%%%%%%%%%%%%%%%%%%%%%%%%%%%%%%%%%%%%%%%%%
\section{Moving Pulse}
%%%%%%%%%%%%%%%%%%%%%%%%%%%%%%%%%%%%%%%%%%%%%%%%%%%%%%%%%%%%%%%%%%%%%%%%%%%%%%%%%%%%%%%%%%%%%%%

So far, we have discussed the situation of a refractive index perturbation $\delta n(t,\f{r})$
appearing in a localized spatial region and then disappearing again.
Now, let us consider the scenario of a moving pulse -- where we first study the case of 
constant velocity $\vau$ 
\bea
\delta n(t,\f{r})=\delta\bar n\,f(\Omega[\f{r}-\vau t]/c)
\,.
\ea
The resulting quantum radiation crucially depends on whether the velocity $\vau$ 
is smaller or larger than the speed of light $c=1/n$ in the medium, see also 
\cite{Faccio-Superluminal}. 
\begin{itemize}
\item 
For a ``sub-luminal'' pulse $v<c$, there is no effect (to lowest order in  $\delta n$). 

This can be understood in the following way: electrodynamics in a 
(homogeneous and isotropic) dielectric medium is formally invariant under 
``Lorentz'' transformations with the speed of light $c_0$ in vacuum being
replaced by the speed of light $c=1/n$ in the medium. 
Thus, after such a ``Lorentz'' boost, the refractive index perturbation 
$\delta n(t,\f{r})$ becomes effectively stationary 
$\delta n(t,\f{r})\to\delta n(\f{r})$ and thus -- as we have seen before -- 
does not produce radiation. 
\item 
For a ``super-luminal'' pulse $v>c$, we get ``quantum Cherenkov'' radiation.

In this case, a suitable ``Lorentz'' boost yields an instantaneous
perturbation $\delta n(t)\times f(\f{r})$.  
\end{itemize}
In the case of a moving pulse (without beginning and end), we cannot compute 
a total emission probability but only the emission probability per unit time 
(as in Fermi's golden rule, for example). 
In the ``super-luminal'' case, the situation after the ``Lorentz'' boost
is analogous to the previous section with $\Omega_1=\Omega_3\to\Omega$ 
and $\Omega_2\to0$ and thus we get 
\bea
\frac{P}{T}=
\ord\left(\Omega\,\delta\bar n^2\right)
%\times 
%\ord\left([v-c]^2\right)
\,.
\ea
Due to phase matching conditions, the photon pairs are predominantly emitted 
in forward direction within an emission angle $\vartheta=\ord(\sqrt{v-c})$.
This angle closes for $v\downarrow c$ and thus the emission probability 
(per unit time) vanishes in this limit -- consistent with the above picture. 
As usual, the typical photon energy scales with $E=\ord(\Omega)$. 

%%%%%%%%%%%%%%%%%%%%%%%%%%%%%%%%%%%%%%%%%%%%%%%%%%%%%%%%%%%%%%%%%%%%%%%%%%%%%%%%%%%%%%%%%%%%%%%
\section{Hawking Radiation}
%%%%%%%%%%%%%%%%%%%%%%%%%%%%%%%%%%%%%%%%%%%%%%%%%%%%%%%%%%%%%%%%%%%%%%%%%%%%%%%%%%%%%%%%%%%%%%%

\begin{figure}[hbt]
\epsfig{file=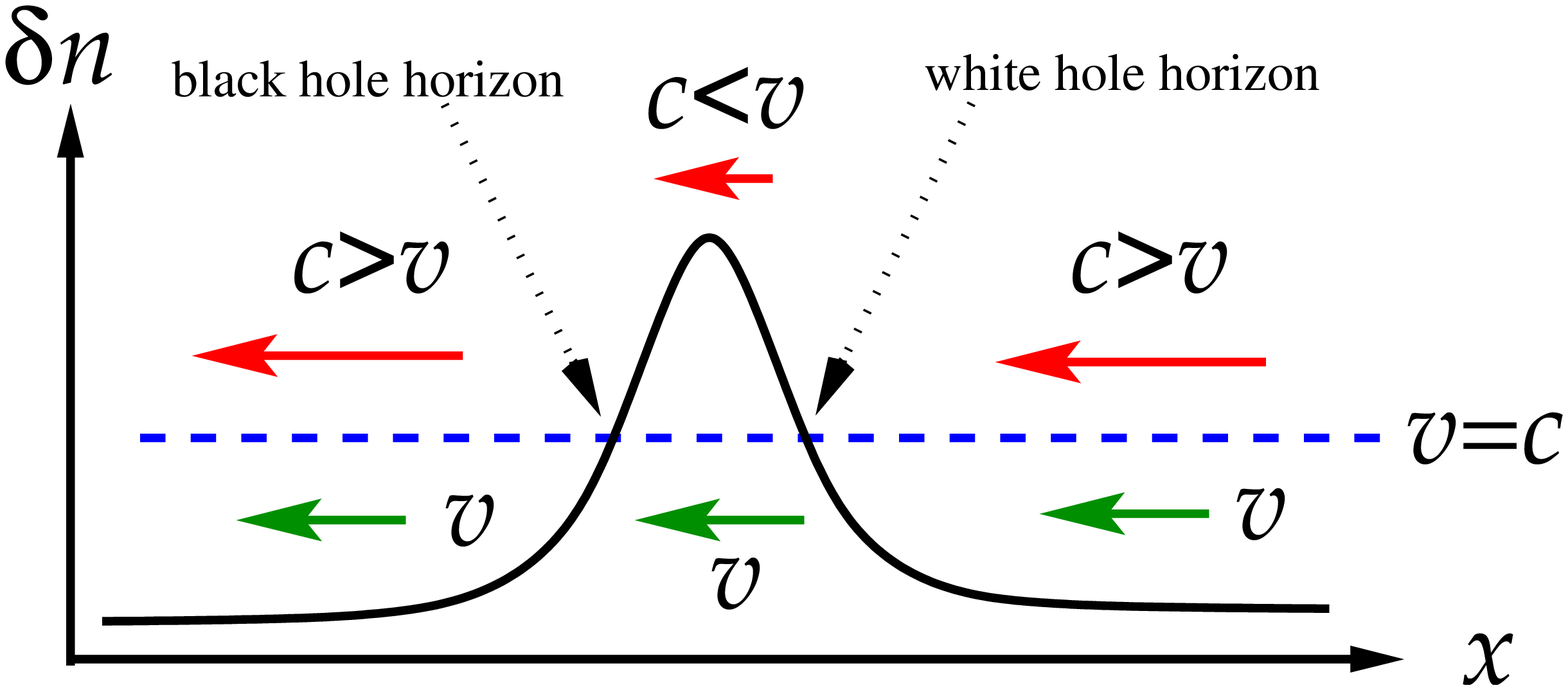,height=.25\textheight}
\caption{Sketch of the pulse creating analogues of black and white hole horizons.}
\label{pulse}
\end{figure}

Now let us go right to the borderline between the ``sub-luminal'' and the 
``super-luminal'' case discussed above and study a pulse which is slower 
than the medium speed of light outside the pulse $c=1/n$ but faster than 
the medium speed of light inside $1/(n+\delta n)$
\bea
\label{zwischen}
\delta n(t,\f{r})=\delta\bar n\,f(\Omega[\f{r}-\vau t]/c)
\,,\quad 
c=\frac1n>v>\frac1{n+\delta\bar n}
\,.
\ea
In this case, we get the analogue of black hole horizon 
(light cannot escape) at the front end of the pulse and 
the analogue of a white hole horizon (light cannot enter)
at its back and, see Fig.~\ref{pulse}. 
The idea of creating analogues of black (or white) holes in the laboratory
and so to experimentally test Hawking radiation \cite{Hawking1,Hawking2}
goes back to Bill Unruh \cite{Dumb}, 
see also \cite{Novello,Volovik,Max-Planck}.
Originally, the analogy was developed for sound waves \cite{Dumb}, 
but later electromagnetic waves were considered as well.
In \cite{Piwnicki,Leonhardt-nature} proposals based on ``slow light''
(i.e., electromagnetically induced transparency) where put forward.
However, these suggestions had various problems, see, e.g., 
\cite{Visser-comment,Piwnicki-reply,Slow}.
A set-up based on moving dielectrics was proposed in \cite{Dielectric},
but there the difficulty was to have the medium flowing faster than 
the speed of light in the medium -- which is typically very fast. 
Later is was realized \cite{wave-guide} that it is not really 
necessary to actually move the medium -- instead a moving pulse 
which changes the local propagation speed can be enough, 
see also \cite{Philbin}.   

%%%%%%%%%%%%%%%%%%%%%%%%%%%%%%%%%%%%%%%%%%%%%%%%%%%%%%%%%%%%%%%%%%%%%%%%%%%%%%%%%%%%%%%%%%%%%%%
%%%%%%%%%%%%%%%%%%%%%%%%%%%%%%%%%%%%%%%%%%%%%%%%%%%%%%%%%%%%%%%%%%%%%%%%%%%%%%%%%%%%%%%%%%%%%%%

The Hawking temperature (in the co-moving frame) is determined by the 
gradient of the refractive index, see, e.g., \cite{Dumb,Dielectric}    
\bea
T_{\rm Hawking}\propto\na\delta n\sim\Omega\,\delta\bar n
\,.
\ea
Since Hawking radiation is thermal, the total emission probability 
(per unit time) scales as 
\bea
\frac{P}{T}=
\ord\left(A\,T_{\rm Hawking}^3\right)
\to
\ord\left(\Omega\,\delta\bar n^3\right)
\,.
\ea
As we may infer from the $\delta\bar n^3$ scaling, this effect is 
beyond lowest order perturbation theory.
In one spatial dimension (such as in an optical fibre, see \cite{Philbin}),  
we would get $P/T\propto\delta\bar n$. 
This shows that this effect is non-perturbative -- since the perturbative
expansion always starts with $\delta\bar n^2$. 
This can be explained by the fact that Hawking radiation occurs due to 
the tearing apart of modes at the black-hole horizon, which takes a time 
duration long compared to $1/(\Omega\,\delta\bar n)$. 
 
An experiment with a pulse satisfying Eq.~(\ref{zwischen}) in some 
frequency range has been done \cite{Faccio-Hawking} and it has been 
claimed that this was the first observation of the analogue of 
Hawking radiation. 
However, comparison with the above estimates casts some doubts at 
these claims, see \cite{comment,reply}.

%%%%%%%%%%%%%%%%%%%%%%%%%%%%%%%%%%%%%%%%%%%%%%%%%%%%%%%%%%%%%%%%%%%%%%%%%%%%%%%%%%%%%%%%%%%%%%%
\section{Non-Inertial Pulse}
%%%%%%%%%%%%%%%%%%%%%%%%%%%%%%%%%%%%%%%%%%%%%%%%%%%%%%%%%%%%%%%%%%%%%%%%%%%%%%%%%%%%%%%%%%%%%%%

After having discussed the case of constant velocity, 
let us turn to non-inertial pulse motion 
\bea
\delta n(t,\f{r})=\delta\bar n\,f(\Omega[\f{r}-\f{r}_P(t)]/c)
\,,
\ea
where $\f{r}_P(t)$ is the trajectory of the pulse.
For simplicity, let us consider the case of approximately uniform 
acceleration $\f{\ddot r}_P\approx\rm const$ for some time. 
In this case, the quantum radiation created by the refractive 
index perturbation can be nicely understood as a signature of the 
Unruh effect.

The Unruh effect \cite{Unruh} describes the striking prediction that a 
uniformly accelerated detector experiences the inertial 
Minkowski vacuum state as thermal bath with Unruh temperature
\bea
T_{\rm Unruh}
=
\frac{\hbar}{2\pi k_{\rm B}c_0}\,a
%=
%\frac{\hbar c}{2\pi k_{\rm B}}
%\,\frac{1}{d_{\rm horizon}}
\,,
\ea
where $a$ is the acceleration. 
Unfortunately, due the the factors $\hbar$ and $c_0$, this temperature 
is quite low for every-day accelerations -- e.g., the earth's gravitational 
acceleration  $a=g=9.81\,{\rm m/s^2}$ corresponds to 
$T_{\rm Unruh}=\ord(10^{-20}\,{\rm K})$. 
This is one of the reasons why this effect has not been directly 
observed yet, see also \cite{Maia,Schaller,Habs} and references therein. 

%%%%%%%%%%%%%%%%%%%%%%%%%%%%%%%%%%%%%%%%%%%%%%%%%%%%%%%%%%%%%%%%%%%%%%%%%%%%%%%%%%%%%%%%%%%%%%%
\begin{figure}[hbt]
\epsfig{file=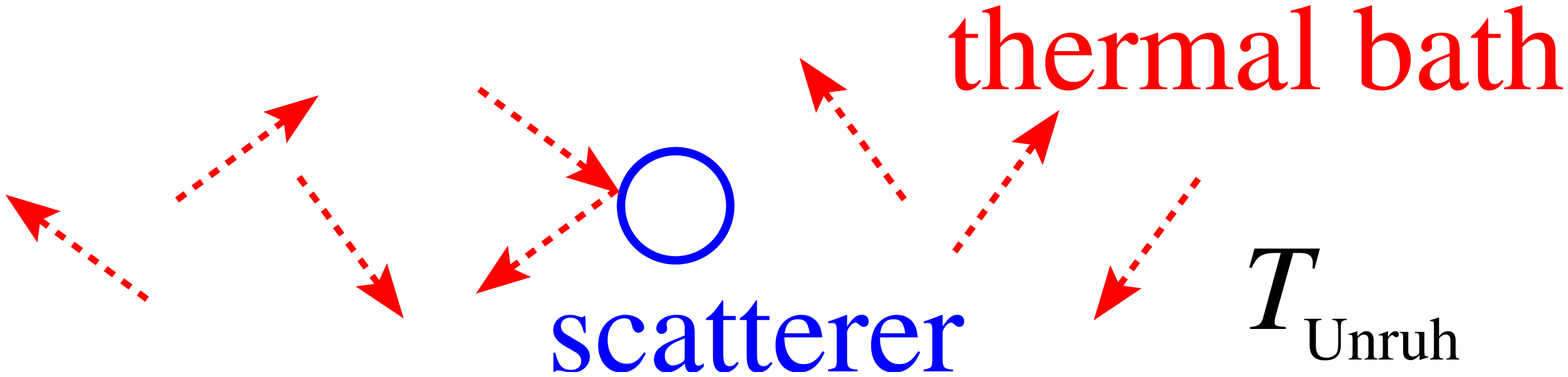,height=.09\textheight}
\hfill
\epsfig{file=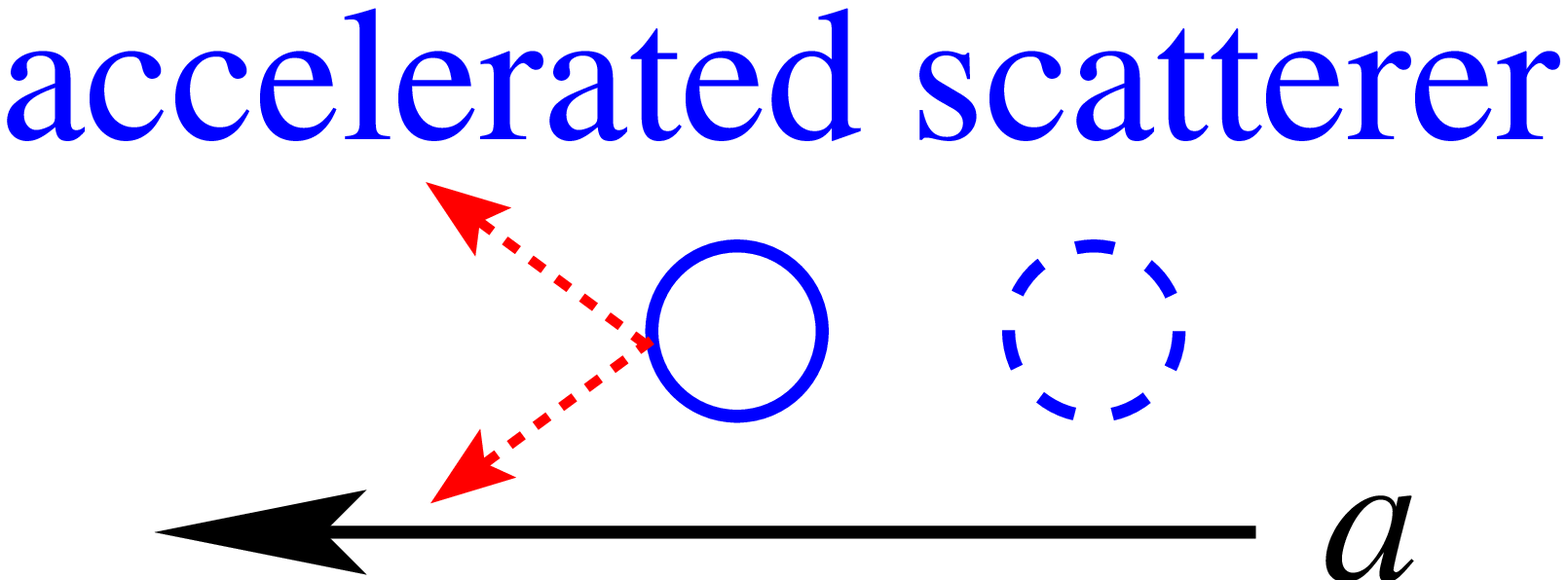,height=.09\textheight}
\caption{Sketch of the scattering of two photons (red arrows) 
in the accelerated frame (left) and translation of this event 
to the inertial frame (right).}
\label{non-inertial}
\end{figure}
%%%%%%%%%%%%%%%%%%%%%%%%%%%%%%%%%%%%%%%%%%%%%%%%%%%%%%%%%%%%%%%%%%%%%%%%%%%%%%%%%%%%%%%%%%%%%%%

However, it should be possible to observe a signature of the Unruh effect
with accelerated refractive index perturbations, for example.
As discussed before, the ground state of a dielectric medium behaves 
analogous to the Minkowski vacuum with the vacuum speed of light $c_0$
being replaced by the medium light velocity $c$.
Thus, in its non-inertial frame, the accelerated refractive index 
perturbation experiences the ground state of the medium as a thermal 
bath. 
Then there is a finite probability that one photon out of this 
thermal bath is scattered by the refractive index perturbation 
into another mode. 
This scattering event in the accelerated frame, when 
translated back into the inertial (laboratory) frame, 
corresponds to the emission of real photon pairs, 
as sketched in Fig.~\ref{non-inertial}.
In agreement with this intuitive picture, 
the emission probability per unit time scale as 
(again using perturbation theory)
\bea
\frac{P}{T}=
\ord\left(\sigma_{\rm scattering}T_{\rm Unruh}^3\right)
\to
\ord\left(\frac{\delta n^2}{\Omega^2}\,\f{\ddot r}^3_P\right)
\,.
\ea
Note that this expression is valid for $\Omega\gg|\f{\ddot r}_P|$ 
only since otherwise the internal width $\sim1/\Omega$ of the pulse 
smears out its trajectory $\f{r}_P(t)$ and thus the picture based 
on the Unruh effect does not apply any-more 
(even though there still would be quantum radiation). 

%%%%%%%%%%%%%%%%%%%%%%%%%%%%%%%%%%%%%%%%%%%%%%%%%%%%%%%%%%%%%%%%%%%%%%%%%%%%%%%%%%%%%%%%%%%%%%%
\section{Dynamical Casimir Effect}
%%%%%%%%%%%%%%%%%%%%%%%%%%%%%%%%%%%%%%%%%%%%%%%%%%%%%%%%%%%%%%%%%%%%%%%%%%%%%%%%%%%%%%%%%%%%%%%

Finally, let us discuss the relation between the quantum radiation given off by these 
refractive index perturbations and the dynamical Casimir effect.
In the original setting, the static Casimir effect \cite{Casimir} describes the 
attraction (or repulsion -- depending on the boundary conditions) of two conducting 
plates at rest in vacuum, which is caused by the distortion of the quantum vacuum 
fluctuations of the electromagnetic field. 
The dynamical Casimir effect then refers to the creation of photon pairs out of QED
vacuum due to the non-inertial motion of one or both mirrors, see, e.g., \cite{Dodonov}.

Replacing the two mirrors by two bodies of dielectric material, one can also get a 
static Casimir attraction.
Thus, the quantum radiation created by the non-inertial motion of a dielectric body 
in vacuum could also be called dynamical Casimir effect.
As we have discussed before, this scenario is formally equivalent to a refractive 
index perturbation moving in a dielectric medium.
Thus, the signatures of the Unruh effect caused by the non-inertial motion of such a 
perturbation as discussed in the previous Section can also be viewed as a 
manifestation\footnote{Note that there is recent experimental evidence \cite{observation} 
for the observation of the dynamical Casimir effect -- though not in vacuum, 
but in a wave-guide (which is analogous to the dielectric medium discussed here).}  
of the dynamical Casimir effect.

%%%%%%%%%%%%%%%%%%%%%%%%%%%%%%%%%%%%%%%%%%%%%%%%%%%%%%%%%%%%%%%%%%%%%%%%%%%%%%%%%%%%%%%%%%%%%%%
\section*{Acknowledgements}
%%%%%%%%%%%%%%%%%%%%%%%%%%%%%%%%%%%%%%%%%%%%%%%%%%%%%%%%%%%%%%%%%%%%%%%%%%%%%%%%%%%%%%%%%%%%%%%

The author acknowledges valuable discussions during the 
{\em II Amazonian Symposium on Physics} in Bel\'em (2011) 
and the 
{\em International Workshop on Dynamical Casimir Effect} in Padova (2011)
as well as financial support by the DFG.  

%%%%%%%%%%%%%%%%%%%%%%%%%%%%%%%%%%%%%%%%%%%%%%%%%%%%%%%%%%%%%%%%%%%%%%%%%%%%%%%%%%%%%%%%%%%%%%%

\end{document}